\documentclass[a4paper,11pt]{article}
\usepackage{pos}

\title{Towards identifying the minimal flavor symmetry behind neutrino oscillations}

\author*[a,b,c]{Zhi-zhong Xing}

\affiliation[a]{Institute of High Energy Physics, Chinese Academy of Sciences, Beijing 100049, China}

\affiliation[b]{School of Physical Sciences, University of Chinese Academy of Sciences, Beijing 100049, China}

\affiliation[c]{Center of High Energy Physics, Peking University, Beijing 100871, China}

\emailAdd{xingzz@ihep.ac.cn}

\abstract{Current neutrino oscillation data indicate that the $3\times 3$ Pontecorvo-Maki-Nakagawa-Sakata 
	matrix $U$ exhibits a $\mu$-$\tau$ flavor interchange symmetry $|U^{}_{\mu i}| = |U^{}_{\tau i}|$ (for 
	$i = 1, 2, 3$) as a good  approximation. In particular, the T2K measurement implies that the maximal 
	neutrino mixing angle $\theta^{}_{23}$ and the CP-violating phase $\delta$ should be close to $\pi/4$ 
	and $-\pi/2$, respectively. Behind these observations lies a minimal flavor symmetry --- the effective 
	Majorana neutrino mass term keeps invariant under the transformations 
	 $\nu^{}_{e \rm L} \to (\nu^{}_{e \rm L})^c$, 
	 $\nu^{}_{\mu \rm L} \to (\nu^{}_{\tau \rm L})^c$, 
	 $\nu^{}_{\tau \rm L} \to (\nu^{}_{\mu \rm L})^c$. Extending this flavor symmetry to the canonical 
	 seesaw mechanism, we find that the $R$-matrix describing the strength of weak charged-current interactions of 
	 heavy Majorana neutrinos satisfies $|R^{}_{\mu i}| = |R^{}_{\tau i}|$ as a consequence of
	 $|U^{}_{\mu i}| = |U^{}_{\tau i}|$. This result can be used to set a new upper bound, 
	 which is about three orders of magnitude more stringent than before, on the flavor mixing 
	 factor associated with the charged-lepton-flavor-violating decay mode $\tau^- \to e^- + \gamma$.}

\FullConference{%
  41st International Conference on High Energy physics - ICHEP2022\\
  6-13 July, 2022\\
  Bologna, Italy
}

\begin{document}
\maketitle

\section{Introduction}

\begin{figure}[h]
	\begin{center}
		\includegraphics[width=3.6in]{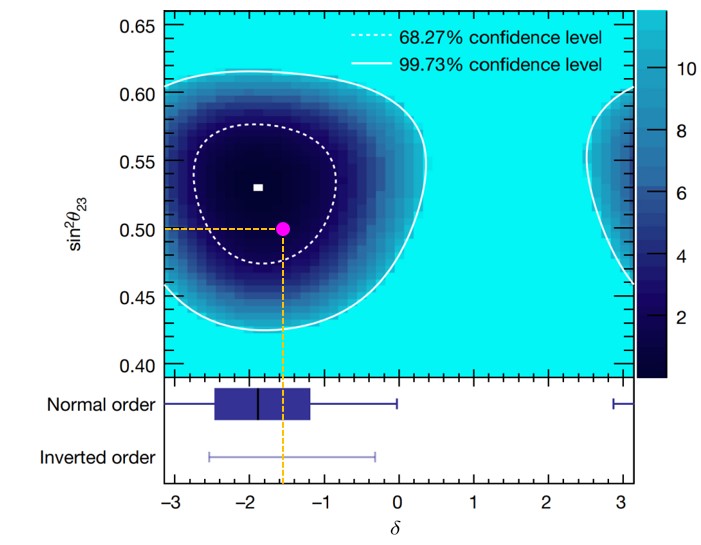}
		\vspace{-.3cm}
		\caption{An illustration of the recent T2K data on the octant of $\theta^{}_{23}$
			and the quadrant of $\delta$~\cite{T2K:2019bcf}, together with their slight 
			deviations from the $\mu$-$\tau$ reflection symmetry predictions 
			$\theta^{}_{23} = \pi/4$ and $\delta = -\pi/2$.}
		\label{Fig:T2K}
	\end{center}
\end{figure}

Current neutrino oscillation data~\cite{ParticleDataGroup:2022pth} allow us to see 
two salient features of the $3\times 3$ Pontecorvo-Maki-Nakagawa-Sakata (PMNS) lepton 
flavor mixing matrix $U$. On the one hand, the observed pattern of $U$ exhibits a 
$\mu$-$\tau$ interchange symmetry $|U^{}_{\mu i}| = |U^{}_{\tau i}|$ (for $i = 1, 2, 3$)
as a very good approximation. On the other hand, the best-fit values of the maximal neutrino 
mixing angle $\theta^{}_{23}$ and the CP-violating phase $\delta$ in the standard 
parametrization of $U$ are close respectively to $\pi/4$ and $-\pi/2$, as illustrated 
in Fig.~\ref{Fig:T2K} according to the recent T2K measurement~\cite{T2K:2019bcf}. 

These two observations motivate us to conjecture that behind $U$ is most likely 
a kind of non-Abelian discrete flavor symmetry group, whose Clebsch-Gordan coefficients may 
help determine some elements of $U$~\cite{Xing:2019vks}. So far a lot of flavor 
symmetries have been proposed and studied. At this stage one is allowed to 
make use of Occam's razor to shave away those complicated models and focus on the minimal flavor
symmetry in the neutrino sector. Such a minimal flavor symmetry should belong to a residual 
flavor symmetry group which is as simple as possible, and its predictions should be as 
close as possible to the experimental data. We argue that the $\mu$-$\tau$ reflection 
symmetry is the most appropriate example in this 
regard~\cite{Harrison:2002et,Xing:2015fdg,Xing:2022uax}: 
the effective Majorana neutrino mass term
\begin{eqnarray}
	-{\cal L}^{}_{\rm mass} = \frac{1}{2} \hspace{0.05cm} \overline{\nu^{}_{\rm L}}
	\hspace{0.05cm} M^{}_\nu (\nu^{}_{\rm L})^c + {\rm h.c.} \; 
\end{eqnarray}
keeps invariant under the transformations
$\nu^{}_{e \rm L} \to (\nu^{}_{e \rm L})^c$,
$\nu^{}_{\mu \rm L} \to (\nu^{}_{\tau \rm L})^c$ and
$\nu^{}_{\tau \rm L} \to (\nu^{}_{\mu \rm L})^c$. The flavor texture of symmetric
$M^{}_\nu$ can therefore be constrained by $M^{}_\nu = {\cal P} M^*_\nu {\cal P}$, where
\begin{eqnarray}
		{\cal P} = {\cal P}^T = {\cal P}^\dagger =
		\left(\begin{matrix} 1 & 0 & 0 \cr
			0 & 0 & 1 \cr 0 & 1 & 0 \cr\end{matrix}\right) \; .
\end{eqnarray}
Then both $\theta^{}_{23} = \pi/4$ and $\delta = \pm \pi/2$ can easily be achieved for
the PMNS matrix $U$ through the diagonalization $U^\dagger M^{}_\nu U^* = D^{}_\nu \equiv
{\rm Diag}\{m^{}_1, m^{}_2, m^{}_3\}$ with $m^{}_i$ being the light neutrino masses.   

Here we are going to identify the same $\mu$-$\tau$ reflection 
symmetry associated with three species of active and sterile neutrinos in the canonical
seesaw mechanism by starting from the very possibility of
$\big|U^{}_{\mu i}\big| = \big|U^{}_{\tau i}\big|$
(for $i=1,2,3$) that is favored by current neutrino oscillation data. 
Considering that $U$ is not exactly unitary and it is intrinsically
correlated with another $3\times 3$ flavor mixing matrix $R$ describing
the strength of weak charged-current interactions of heavy Majorana neutrinos,
we show that $\big|U^{}_{\mu i}\big| = \big|U^{}_{\tau i}\big|$ 
gives rise to a novel prediction $\big|R^{}_{\mu i}\big| = \big|R^{}_{\tau i}\big|$
(for $i = 1, 2, 3$). We prove that behind these two sets of equalities
and the preliminary experimental evidence for leptonic CP violation~\cite{T2K:2019bcf}
lies an expected minimal flavor symmetry; namely, the overall neutrino mass term keeps
invariant when the left-handed neutrino fields transform as
$\nu^{}_{e \rm L} \to (\nu^{}_{e \rm L})^c$,
$\nu^{}_{\mu \rm L} \to (\nu^{}_{\tau \rm L})^c$,
$\nu^{}_{\tau \rm L} \to (\nu^{}_{\mu \rm L})^c$ and the right-handed
neutrino fields undergo an arbitrary unitary transformation. We find that
this novel result can be used to set a new upper limit, which is about three orders of magnitude 
more stringent than before, on the flavor mixing factor associated with the 
charged-lepton-flavor-violating decay mode $\tau^- \to e^- + \gamma$.

\section{The canonical seesaw mechanism}

To explain why the masses of three active neutrinos are tiny, a minimal
extension of the standard model (SM) of particle physics is to add three right-handed 
neutrino fields $N^{}_{\alpha \rm R}$ (for $\alpha = e, \mu, \tau$) and allow for 
lepton number violation. In this case the neutrino mass terms can be written as
\begin{eqnarray}
	-{\cal L}^{}_\nu = \overline{\ell^{}_{\rm L}} \hspace{0.05cm} Y^{}_\nu
	\widetilde{H} N^{}_{\rm R} + \frac{1}{2} \hspace{0.05cm} \overline{(N^{}_{\rm R})^c}
	\hspace{0.05cm} M^{}_{\rm R} N^{}_{\rm R} + {\rm h.c.} \; ,
\end{eqnarray}
where $\ell^{}_{\rm L}$ denotes the $\rm SU(2)^{}_{\rm L}$ doublet of the left-handed
lepton fields, $\widetilde{H} \equiv {\rm i} \sigma^{}_2 H^*$ with
$H$ being the Higgs doublet and $\sigma^{}_2$ being the second Pauli matrix,
$N^{}_{\rm R} = (N^{}_{e \rm R} , N^{}_{\mu \rm R} , N^{}_{\tau \rm R})^T$ is the column
vector of three right-handed neutrino fields which are the $\rm SU(2)^{}_{\rm L}$
singlets, $(N^{}_{\rm R})^c \equiv {\cal C} \overline{N^{}_{\rm R}}^T$ denotes the
charge conjugation of $N^{}_{\rm R}$. After spontaneous gauge symmetry breaking,
Eq.~(3) becomes
\begin{eqnarray}
	-{\cal L}^\prime_\nu = \frac{1}{2} \hspace{0.05cm} \overline{\left[
		\nu^{}_{\rm L} \hspace{0.2cm} (N^{}_{\rm R})^c\right]}
	\left(\begin{matrix} {\bf 0} & M^{}_{\rm D} \cr
		M^T_{\rm D} & M^{}_{\rm R} \end{matrix}\right)
	\left[\begin{matrix} (\nu^{}_{\rm L})^c \cr N^{}_{\rm R} \end{matrix}
	\right] + {\rm h.c.} \; ,
\end{eqnarray}
where $\nu^{}_{\rm L} = (\nu^{}_{e \rm L}, \nu^{}_{\mu \rm L}, \nu^{}_{\tau \rm L})^T$
denotes the column vector of three left-handed neutrino fields,
$M^{}_{\rm D} \equiv Y^{}_\nu \langle H\rangle$ with $\langle H\rangle$ being the
vacuum expectation value of the Higgs field. The overall symmetric $6\times 6$
neutrino mass matrix in Eq.~(4) can be diagonalized by the
unitary transformation
\begin{eqnarray}
	\left( \begin{matrix} U & R \cr S & Q \cr \end{matrix}
	\right)^{\hspace{-0.05cm} \dagger} \left ( \begin{matrix} {\bf 0} & M^{}_{\rm D}
		\cr M^{T}_{\rm D} & M^{}_{\rm R} \cr \end{matrix} \right ) \left(
	\begin{matrix} U & R \cr S & Q \cr \end{matrix} \right)^{\hspace{-0.05cm} *}
	= \left( \begin{matrix} D^{}_\nu & {\bf 0} \cr {\bf 0} &
		D^{}_N \cr \end{matrix} \right) \; ,
\end{eqnarray}
where $D^{}_N \equiv {\rm Diag}\big\{M^{}_1, M^{}_2, M^{}_3 \big\}$ with 
$M^{}_i$ being the heavy Majorana neutrino masses. The $3\times 3$ submatrices $U$, 
$R$, $S$ and $Q$ satisfy the unitarity conditions:
\begin{eqnarray}
	U U^\dagger + RR^\dagger = SS^\dagger + Q Q^\dagger
	\hspace{-0.2cm} & = & \hspace{-0.2cm} I \; ,
	\nonumber \\
	U^\dagger U + S^\dagger S = R^\dagger R + Q^\dagger Q
	\hspace{-0.2cm} & = & \hspace{-0.2cm} I \; ,
	\nonumber \\
	U S^\dagger + R Q^\dagger = U^\dagger R + S^\dagger Q
	\hspace{-0.2cm} & = & \hspace{-0.2cm} {\bf 0} \; ; \hspace{0.5cm}
\end{eqnarray}
and the exact {\it seesaw} formula that characterizes a kind of balance between the
light and heavy neutrino sectors can also be obtained from Eq.~(5):
\begin{eqnarray}
	U D^{}_\nu U^T + R D^{}_N R^T = {\bf 0} \; .
\end{eqnarray}
The smallness of $m^{}_i$ is therefore ascribed to the highly suppressed
magnitude of $R$ which signifies the largeness of $M^{}_i$ with respect to the electroweak
scale. In this seesaw framework $\nu^{}_{\alpha}$
(for $\alpha = e, \mu, \tau$) can be expressed as a linear combination of
the mass eigenstates of three active (light) neutrinos and three sterile (heavy)
neutrinos (i.e., $\nu^{}_{i} = (\nu^{}_i)^c$ and
$N^{}_{i} = (N^{}_i)^c$ for $i = 1, 2, 3$). The standard weak charged-current
interactions of these six Majorana neutrinos is given by
\begin{eqnarray}
	-{\cal L}^{}_{\rm cc} = \frac{g}{\sqrt{2}} \hspace{0.1cm}
	\overline{\left(e \hspace{0.2cm} \mu \hspace{0.2cm}
		\tau \right)^{}_{\rm L}} \hspace{0.1cm} \gamma^\mu
	\left[ U \left( \begin{matrix} \nu^{}_{1} \cr \nu^{}_{2} \cr
		\nu^{}_{3} \cr\end{matrix} \right)^{}_{\rm L} + R \left(
	\begin{matrix} N^{}_{1} \cr N^{}_{2} \cr N^{}_{3}
		\cr\end{matrix}\right)^{}_{\rm L} \right] W^-_\mu + {\rm h.c.} \; .
\end{eqnarray}
So $U$ describes flavor mixing and CP violation of three active neutrinos in neutrino 
oscillations, and $R$ measures the strength of weak charged-current interactions of
three heavy neutrinos in precision collider physics and thermal leptogenesis.
Note that $U$ must be non-unitary, but its deviation from unitarity is at most of 
${\cal O}(1\%)$ and hence is insensitive to all the present experiments.

\section{The minimal flavor symmetry}

As pointed out in Ref.~\cite{Xing:2022oob}, the possibilities of 
$\big|U^{}_{\mu i}\big| = \big|U^{}_{\tau i}\big|$ and $\delta =\pm \pi/2$
inspire us to conjecture how $U^{}_{\mu i}$ is directly related
to $U^*_{\tau i}$ under a simple flavor symmetry. In the same flavor symmetry 
$\big|U^{}_{e i}\big|$ should keep unchanged. We find that 
$U = {\cal P} U^* \zeta$ with $\zeta = {\rm Diag}\{\eta^{}_1, \eta^{}_2, \eta^{}_3\}$
and $\eta^{}_i = \pm 1$, where $\cal P$ has been given in
Eq.~(2), satisfies our requirements and allows the PMNS matrix elements 
$U^{}_{\alpha i}$ (for $\alpha = e, \mu, \tau$) to transform together.
Inserting $U = {\cal P} U^* \zeta$ into Eq.~(7) and taking the complex conjugate 
for the whole equation, we arrive at 
\begin{eqnarray}
	U D^{}_\nu U^T + {\cal P} R^* D^{}_N ({\cal P} R^*)^T = {\bf 0} \; .
\end{eqnarray}
Comparing this equation with Eq.~(7), we immediately get 
$R = {\cal P} R^* \zeta^\prime$ with 
$\zeta^\prime = {\rm Diag}\{\eta^{\prime}_1, \eta^{\prime}_2, \eta^{\prime}_3\}$
and $\eta^{\prime}_i = \pm 1$. This result leads us to a novel and rephasing-invariant 
prediction $\big|R^{}_{\mu i}\big| = \big|R^{}_{\tau i}\big|$, which is actually a 
natural consequence of $\big|U^{}_{\mu i}\big| = \big|U^{}_{\tau i}\big|$ in
the canonical seesaw mechanism. 

Substituting $U = {\cal P} U^* \zeta$ and $R = {\cal P} R^* \zeta^\prime$ 
into Eq.~(6), we find $S = {\cal T} S^* \zeta$ and $Q = {\cal T} Q^* \zeta^\prime$ 
with $\cal T$ being an arbitrary unitary matrix. Let us proceed to insert all these
four relations into Eq.~(5) and then take the complex conjugate for the whole 
equation. We are therefore left with 
\begin{eqnarray}
	\left( \begin{matrix} U & R \cr S & Q \cr \end{matrix}
	\right)^{\hspace{-0.05cm} \dagger} \left ( \begin{matrix} {\bf 0}
		& {\cal P} M^{*}_{\rm D} {\cal T}
		\cr {\cal T}^T M^{\dagger}_{\rm D} {\cal P}
		& {\cal T}^T M^{*}_{\rm R} {\cal T} \cr \end{matrix} \right ) \left(
	\begin{matrix} U & R \cr S & Q \cr \end{matrix} \right)^{\hspace{-0.05cm} *}
	= \left( \begin{matrix} D^{}_\nu & {\bf 0} \cr {\bf 0} &
		D^{}_N \cr \end{matrix} \right) \; .
\end{eqnarray}
A comparison between Eqs.~(5) and (10) yields 
\begin{eqnarray}
	M^{}_{\rm D} = {\cal P} M^{*}_{\rm D} {\cal T} \; , \quad
	M^{}_{\rm R} = {\cal T}^T M^{*}_{\rm R} {\cal T} \; . 
\end{eqnarray}
Inserting Eq.~(11) into Eq.~(4), the overall neutrino mass term ${\cal L}^\prime_\nu$ 
reads as 
\begin{eqnarray}
	-{\cal L}^{\prime\prime}_\nu =
	\frac{1}{2} \hspace{0.05cm} \overline{\left[
		{\cal P} (\nu^{}_{\rm L})^c \hspace{0.37cm} {\cal T} N^{}_{\rm R}\right]}
	\left(\begin{matrix} {\bf 0} & M^{}_{\rm D} \cr
		M^T_{\rm D} & M^{}_{\rm R} \end{matrix}\right)
	\left[\begin{matrix} {\cal P} \nu^{}_{\rm L} \cr
		{\cal T}^* (N^{}_{\rm R})^c \end{matrix}
	\right] + {\rm h.c.} \; .
\end{eqnarray}
Comparing between Eqs.~(4) and (12), 
we find that ${\cal L}^{\prime\prime}_\nu = {\cal L}^\prime_\nu$ holds under the 
transformations~\cite{Xing:2022oob}
\begin{eqnarray}
	\nu^{}_{\rm L} \to {\cal P} (\nu^{}_{\rm L})^c \; , \quad
	N^{}_{\rm R} \to {\cal T}^* (N^{}_{\rm R})^c \; ,
\end{eqnarray}
no matter what specific form of $\cal T$ is taken.  

If the heavy degrees of freedom in the canonical seesaw mechanism are integrated out,
one may obtain the effective Majorana neutrino mass term in Eq.~(1). The 
corresponding effective mass matrix $M^{}_\nu$ can be given by the approximate
seesaw relation $M^{}_\nu \simeq - M^{}_{\rm D} M^{-1}_{\rm R} M^T_{\rm D}$. 
Inserting Eq.~(11) into this formula and take the complex conjugate, we arrive
at $M^{}_\nu = {\cal P} M^*_\nu {\cal P}$ (namely, $M^{}_\nu$ respects the 
$\mu$-$\tau$ reflection symmetry although it originates from the seesaw mechanism). 
\begin{figure}[t]
	\begin{center}
		\includegraphics[width=5.93in]{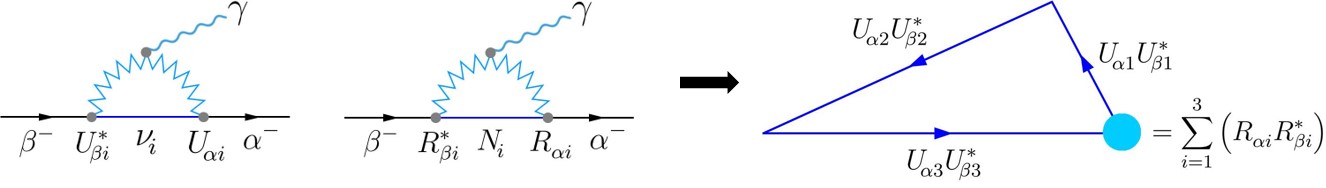}
		\vspace{-.7cm}
		\caption{The charged-lepton-flavor-violating $\beta^- \to \alpha^- + \gamma$ decays 
			mediated by the light Majorana neutrinos $\nu^{}_i$ and the heavy 
			Majorana neutrinos $N^{}_i$ (for $i = 1, 2, 3$), where $\alpha$ and $\beta$ 
			run over the $e$, $\mu$ and $\tau$ flavors and $m^{}_\beta > m^{}_\alpha$ holds.
			The related unitarity hexagons, reduced as the effective $U$-dominated unitarity 
			triangles corrected by the effective $R$-dominated vertices, are also shown.}
		\label{Fig:cLFV}
	\end{center}
\end{figure}

\section{Charged lepton flavor violation}

Now we show that the $\mu$-$\tau$ reflection symmetry in the canonical seesaw
framework can help constrain the unitarity of $U$ through the 
charged-lepton-flavor-violating decay modes $\beta^- \to \alpha^- + \gamma$ 
(for $\alpha, \beta = e, \mu, \tau$ and $m^{}_\beta > m^{}_\alpha$)
as illustrated in Fig.~\ref{Fig:cLFV}, which are mediated by both the light Majorana 
neutrino $\nu^{}_i$ and the heavy Majorana neutrino $N^{}_i$ (for $i = 1, 2, 3$).
For this purpose, we are more interested in the following ratios in the natural 
case of $m^{}_i \ll M^{}_W \ll M^{}_i$~\cite{Xing:2020ivm,Zhang:2021tsq}: 
\begin{eqnarray}
\xi^{}_{\alpha\beta} \hspace{-0.2cm} & \equiv & \hspace{-0.2cm}
\frac{\Gamma\big(\beta^- \to \alpha^- + \gamma\big)}
{\Gamma\big(\beta^- \to \alpha^- + \overline{\nu}^{}_\alpha
+ \nu^{}_\beta\big)} \;
\nonumber \\
\hspace{-0.2cm} & \simeq & \hspace{-0.2cm}
\frac{3\alpha^{}_{\rm em}}{2\pi} \left|
\sum^{3}_{i=1} U^{}_{\alpha i} U^\ast_{\beta i} \left(-\frac{5}{6}
+ \frac{1}{4}\cdot \frac{m^2_i}{M^2_W}\right) -\frac{1}{3} \sum^{3}_{i=1}
R^{}_{\alpha i} R^\ast_{\beta i} \right|^2
\nonumber \\
\hspace{-0.2cm} & \simeq & \hspace{-0.2cm}
\frac{3\alpha^{}_{\rm em}}{8\pi} \left|
\sum^{3}_{i=1} U^{}_{\alpha i} U^\ast_{\beta i} \right|^2 = 
\frac{3\alpha^{}_{\rm em}}{8\pi} \left|
\sum^{3}_{i=1} R^{}_{\alpha i} R^\ast_{\beta i} \right|^2 \; ,	
\end{eqnarray}
where $\alpha^{}_{\rm em}$ is the fine structure constant of electromagnetic interactions,
and $U U^\dagger + R R^\dagger = I$ has been used. Then we obtain 
\begin{eqnarray}
	\left| \sum^{3}_{i=1} R^{}_{\alpha i} R^\ast_{\beta i} \right| =
	\left| \sum^{3}_{i=1} U^{}_{\alpha i} U^\ast_{\beta i} \right| =
	\sqrt{\frac{8\pi}{3 \alpha^{}_{\rm em}} \xi^{}_{\alpha\beta}}
	\simeq 33.88 \sqrt{\xi^{}_{\alpha\beta}} \; ,
\end{eqnarray}
where $\alpha^{}_{\rm em} \simeq 1/137$ has been input. Taking account of current
experimental upper bounds on the branching ratios of 
$\beta^- \to \alpha^- + \gamma$~\cite{ParticleDataGroup:2022pth}, we have
$\xi^{}_{e\mu} < 4.20 \times 10^{-13}$, $\xi^{}_{e\tau} < 1.85 \times 10^{-7}$ and
$\xi^{}_{\mu\tau} < 2.42 \times 10^{-7}$. So Eq.~(15) give the constraints 
\begin{eqnarray}
		\left| \sum^{3}_{i=1} R^{}_{e i} \hspace{0.025cm} R^\ast_{\mu i} \right| <
	2.20 \times 10^{-5} \; , \quad
		\left| \sum^{3}_{i=1} R^{}_{e i} \hspace{0.045cm} R^\ast_{\tau i} \right| <
	1.46 \times 10^{-2} \; , \quad
		\left| \sum^{3}_{i=1} R^{}_{\mu i} R^\ast_{\tau i} \right| <
	1.66 \times 10^{-2} \; .
\end{eqnarray} 
These results mean that the effective $R$-dominated vertices of the effective 
$U$-dominated unitarity triangles in Fig.~\ref{Fig:cLFV} are really small, at
most at the ${\cal O}(1\%)$ level.  

As we have shown in section 3, $R$ respects the $\mu$-$\tau$ reflection symmetry
as $U$ does. In this case we can achieve a new upper bound on the flavor mixing 
factor associated with $\tau^- \to e^- + \gamma$, which is 
about three orders of magnitude more stringent than that obtained in Eq.~(16):
\begin{eqnarray}
	\left| \sum^{3}_{i=1} R^{}_{e i} R^\ast_{\tau i} \right| =
	\left| \sum^{3}_{i=1} R^{}_{e i} R^\ast_{\mu i} \right| <
	2.20 \times 10^{-5} \; .
\end{eqnarray}
Some other important applications of the $\mu$-$\tau$ reflection symmetry in
neutrino phenomenology have recently been reviewed in Refs.~\cite{Xing:2015fdg,Xing:2022uax}. 
A further study of its implications is well worth the wait. 

This work was supported in part by the NSFC under grants No. 12075254 and No. 11835013.

\end{document}